\documentclass[sigconf]{acmart}

\AtBeginDocument{%
  \providecommand\BibTeX{{%
    Bib\TeX}}}

\usepackage{algorithmic}
\usepackage{graphicx}
\usepackage{textcomp}
\usepackage{xcolor}
\usepackage{multirow}
\usepackage{multicol}
\usepackage{tikz}
\usepackage{multirow}
\usepackage{wrapfig}  
\usepackage{caption}
\usepackage{subcaption}
\usepackage{booktabs}
\usepackage[ruled, vlined, linesnumbered]{algorithm2e}
\usepackage{bm}
\usepackage{pifont}
\def\BibTeX{{\rm B\kern-.05em{\sc i\kern-.025em b}\kern-.08em
    T\kern-.1667em\lower.7ex\hbox{E}\kern-.125emX}}
    
\setcopyright{acmlicensed}
\copyrightyear{2018}
\acmYear{2018}
\acmDOI{XXXXXXX.XXXXXXX}

\acmConference[Conference acronym 'XX]{Make sure to enter the correct
  conference title from your rights confirmation emai}{June 03--05,
  2018}{Woodstock, NY}
\acmISBN{978-1-4503-XXXX-X/18/06}

\copyrightyear{2024}
\acmYear{2024}
\setcopyright{acmlicensed}\acmConference[DAC '24]{61st ACM/IEEE Design Automation Conference}{June 23--27, 2024}{San Francisco, CA, USA}
\acmBooktitle{61st ACM/IEEE Design Automation Conference (DAC '24), June 23--27, 2024, San Francisco, CA, USA}
\acmDOI{10.1145/3649329.3655679}
\acmISBN{979-8-4007-0601-1/24/06}
\begin{document}

\title{PIVOT- Input-aware Path Selection for Energy-efficient ViT Inference}

\author{Abhishek Moitra, Abhiroop Bhattacharjee and Priyadarshini Panda\\
Yale University, New Haven, CT, 06511, USA}

\begin{abstract}

The attention module in vision transformers(ViTs) performs intricate spatial correlations, contributing significantly to accuracy and delay. It is thereby important to modulate the number of attentions according to the input feature complexity for optimal delay-accuracy tradeoffs. To this end, we propose PIVOT - a co-optimization framework which selectively performs attention skipping based on the input difficulty. For this, PIVOT employs a hardware-in-loop co-search to obtain optimal attention skip configurations. Evaluations on the ZCU102 MPSoC FPGA show that PIVOT achieves 2.7$\times$ lower EDP at 0.2\% accuracy reduction compared to LVViT-S ViT. PIVOT also achieves 1.3\% and 1.8$\times$ higher accuracy and throughput than prior works on traditional CPUs and GPUs. The PIVOT project can be found at \href{https://github.com/Intelligent-Computing-Lab-Yale/PIVOT}{\textcolor{blue}{\underline{{this Github link.}}}}

\end{abstract}


\maketitle 

\section{Introduction}
\label{sec:intro}

\begin{figure}[t]
    \centering
    \includegraphics[width=\linewidth]{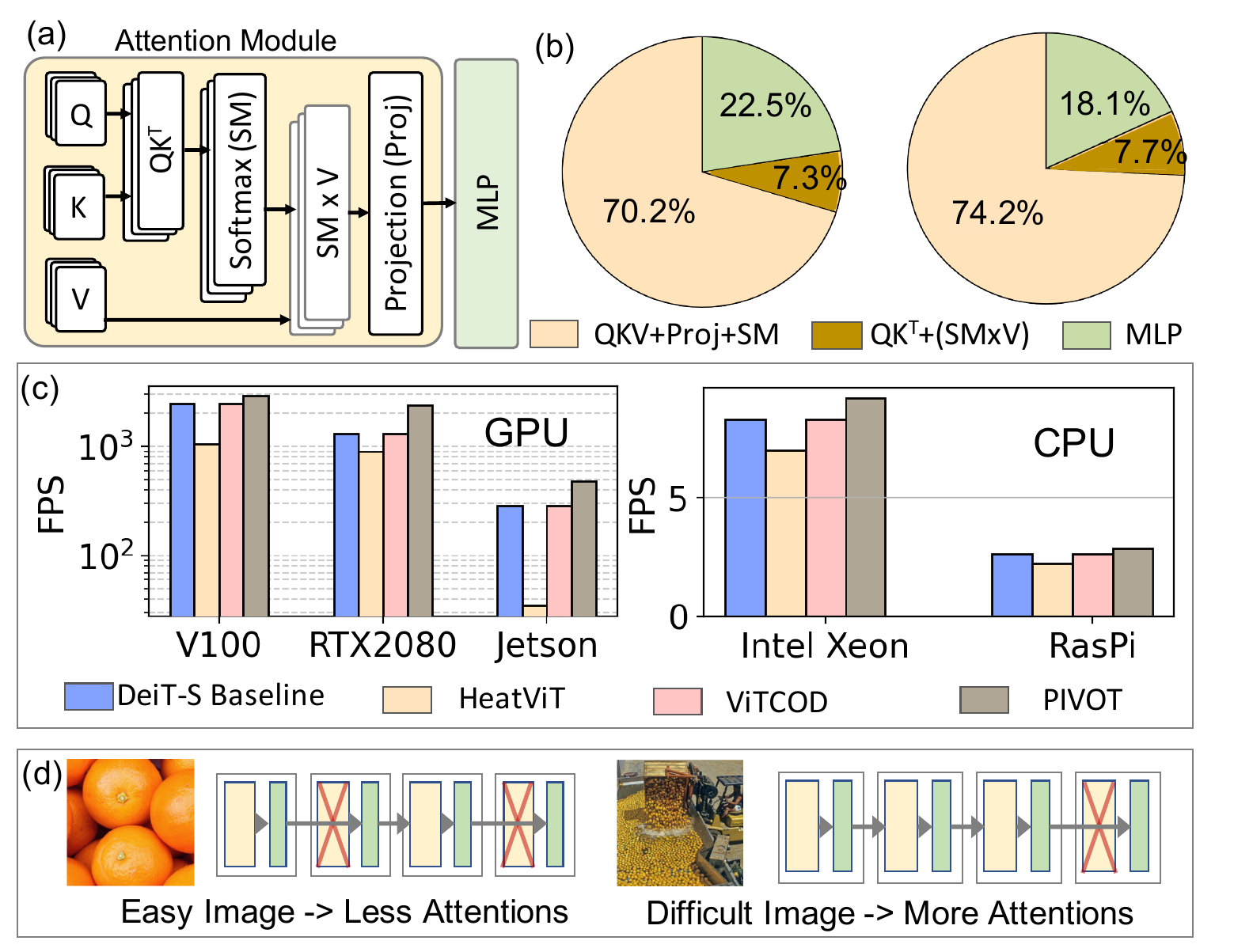}
    \caption{(a) Figure showing the encoder architecture of a vision transformer. Q-Query, K-Key and V-Value. (b) Delay distribution across different ViT modules for DeiT-S (left) and LVViT-S (right) ViTs. Note, Attention delay is QKV+SM+QK$^T$+(SMxV)+Proj. (c) Throughput of PIVOT compared with DeiT-S Baseline (a standard DeiT-S \cite{touvron2021training} ViT), prior token pruning (HeatViT \cite{dong2023heatvit}) and attention sparsification (ViTCOD \cite{you2023vitcod}) techniques implemented on GPUs- Nvidia V100, RTX2080ti, Jetson Orin Nano and CPUs- Intel Xeon and Raspberry Pi 4. (d) PIVOT's input difficulty-aware inference } 
    \label{fig:flop_red_latency_ov}
\end{figure}

Vision Transformers (ViT) have demonstrated remarkable accuracy in large-scale image classification tasks \cite{dehghani2023scaling, han2022survey, dosovitskiy2020image}. The success of ViTs can be attributed to the attention module shown in Fig. \ref{fig:flop_red_latency_ov}a which utilizes the self-attention mechanism to perform sophisticated spatial correlation operations \cite{han2022survey}. However, the attention module, involves computationally intensive operations, including matrix multiplications and non-linear functions like softmax \cite{han2022survey,dehghani2023scaling}. Hence, as seen in Fig. \ref{fig:flop_red_latency_ov}b, the attention module (QKV+QK$^T$+SM+(SMxV)+Proj combined) contributes 77.5\% to 81.9\% of the total ViT inference delay. 

Recently, there have been several ViT inference optimization works that focus on reducing the attention delay overhead. These mainly fall under two categories 1) Attention sparsification \cite{kim2021rethinking, you2023vitcod} 2) Token pruning techniques \cite{dong2023heatvit, rao2021dynamicvit, wang2021spatten}. Attention sparsification techniques exploit the sparsity in the QK$^T$ and (SMxV) layers \cite{kim2021rethinking, you2023vitcod} (Fig. \ref{fig:flop_red_latency_ov}a). In \cite{kim2021rethinking}, the authors algorithmically investigate the effect of structured sparsity in the attention heads on ViT accuracy. 
In a more recent work \cite{you2023vitcod}, the authors propose an accelerator co-design framework that performs sparse-dense attention decomposition and develop a sparse accelerator to exploit the attention sparsity. The objective of token pruning is to selectively reduce the number of tokens in the ViT. In \cite{rao2021dynamicvit, wang2021spatten}, the authors use predictor networks to compute the global-local token importance to eliminate redundant tokens. In HeatViT \cite{dong2023heatvit}, the authors use predictor networks to score the token importance based on the information in each attention head. Along with the predictor networks, the authors use a token packaging technique wherein unimportant tokens are combined into one token to maintain a good accuracy-efficiency tradeoff. Although, attention sparsification and token pruning works \cite{kim2021rethinking, you2023vitcod, rao2021dynamicvit, dong2023heatvit} achieve good accuracy at reduced computation, they have two major problems. Firstly, the portion of delay optimized by these works is small. For example, attention sparsification works are only able to optimize 7.3-7.7\% of the overall delay since they target the QK$^T$ and (SMxV) layers as shown in Fig. \ref{fig:flop_red_latency_ov}b. The second
problem is that attention sparsification and token pruning approaches require nuanced hardware support to achieve optimal efficiency. For example, attention sparsification works require sparse matrix multiplication hardware to fully exploit sparse computations. Similarly, token pruning works require custom hardware design to efficiently implement the token score predictor modules.  
Thus, as shown in Fig. \ref{fig:flop_red_latency_ov}c, when implemented on general purpose platforms (GPPs) such as CPUs and GPUs, they do not achieve any inference delay benefits and, in fact, result in lower throughput compared to a dense baseline.

Another missing consideration in prior ViT optimization literature is the input difficulty awareness. Interestingly, different images have different feature complexity. For example, an easy image will contain simple, low-level features compared to a difficult image with intricate feature representations \cite{wu2018blockdrop}. Since attention modules are responsible for capturing different levels of feature representations in the image, it is therefore imperative to modulate the number of attentions in a ViT according to the input difficulty (Fig. \ref{fig:flop_red_latency_ov}d). Modulating the number of attentions according to input difficulty will ensure minimal attention activation to achieve high accuracy at low inference delay. There have been several input difficulty-aware network optimization works in the CNN literature \cite{panda2016conditional, wu2018blockdrop, bhattacharjee2022mime}. However, there are no works that analyze the co-dependency between the number of attentions and input difficulty from the perspective of accuracy and ViT inference delay.

To this end, we propose PIVOT, a hardware-algorithm co-design framework that modulates the number of attentions in the ViT according to the input difficulty. The goal of PIVOT is to achieve high classification accuracy by using the minimum number of attentions in the ViT. As shown in Fig. \ref{fig:flop_red_latency_ov}d during inference, PIVOT uses two kinds of ViTs - 1) Low Effort and 2) High Effort ViT. The low effort ViT entails more attention skips compared to the high effort and classifies the easy images. While the high effort ViT is used for classifying the difficult images. An iterative hardware-in-the-loop co-search is applied to obtain the optimal low and high effort ViTs according to the user-provided delay constraints. For evaluation, we implement PIVOT on various GPPs such as CPUs and GPUs. Additionally, we also evaluate PIVOT on Xilinx ZCU102-implemented systolic array accelerator \cite{samajdar2018scale}. Unlike token pruning and attention sparsification works, PIVOT does not require any application-specific hardware and can achieve 1.3$\times$-2$\times$ higher throughput than baseline across various GPPs as shown in Fig. \ref{fig:flop_red_latency_ov}c. 

In summary, the key contributions of our work are:
\begin{enumerate}
    \item We propose PIVOT- a hardware-algorithm co-optimization framework that leverages input difficulty-aware attention skipping in ViTs to overcome the high inference delay overhead of the attention module. During attention optimization, PIVOT uses PIVOT-Sim, a cycle-accurate simulator for ViT implemented on a Xilinx ZCU102 FPGA-based systolic array accelerator. PIVOT-Sim will be made open-source and can benchmark different state-of-the-art ViTs.
    
    \item Using PIVOT-Sim, we find that PIVOT achieves 1.73$\times$ (2.7$\times$) lower energy-delay-product (EDP) at merely 0.4\% (0.2\%) accuracy reduction compared to DeiT-S \cite{touvron2021training} (LVViT-S \cite{jiang2021all}) baselines. End-to-end  evaluations using PIVOT-Sim show that PIVOT is able to achieve more than 1.7$\times$ energy reduction across different resources in the Xilinx ZCU102 FPGA such as the ZynQ MPSoC PS, systolic array, on-chip buffers, and communication/memory controller circuits. 

    \item Through extensive experiments we show the overheads introduced by prior ViT co-optimization works \cite{dong2023heatvit, you2023vitcod} when implemented on GPPs such as GPUs and CPUs. As PIVOT does not require nuanced hardware support, when implemented on GPPs, it achieves 1.8$\times$ higher throughput at 0.4-1.3\% higher accuracy compared to prior works. 
    

\end{enumerate}

\section{Background on Vision Transformer}
\label{sec:motivation}


A Vision Transformer (ViT) comprises multiple cascaded encoders, and each encoder follows the architecture depicted in Fig. \ref{fig:flop_red_latency_ov}a. In each encoder, the inputs of dimensions $t\times d$ undergo QKV operations wherein, weights $W_Q$, $W_K$ and $W_V$ are multiplied with the input to generate the Query (Q), Key (K) and Value (V) matrices. The attention module uses the multi-head self-attention (MHSA) mechanism, that captures close relationships between different image features \cite{touvron2021training, jiang2021all, rao2021dynamicvit}. For this, the Q, K and V outputs are partitioned into multiple smaller attention heads ($Q_i$, $K_i$, $V_i$), where $i$ denotes a head of MHSA.

The attention is computed using Equation \ref{eq:attention_eq}. In each head matrix multiplications between $Q_i$, $K_i^T$ ($QK^T$) is performed followed by the softmax (SM) and matrix multiplication with $V_i$ (SM$\times$V) operations \cite{stevens2021softermax}. The softmax is computed using Equation \ref{eq:softmax}. 


\begin{equation}
Attention(Q_i,K_i,V_i) = {Softmax(\frac{Q_iK_i^T}{\sqrt{d}})}V_i,
\label{eq:attention_eq}
\end{equation}
\begin{equation}
    Softmax(x_i) = \frac{e^{x_i-x_{max}}}{\sum_i {e^{x_i-x_{max}}}}.
    \label{eq:softmax}
\end{equation}

Next, the attention outputs are concatenated resulting in a $t\times d$ output attention matrix. Following this, the projection and MLP layers project the information into a higher dimension feature space. Each encoder outputs a $t\times d$ vector that is forwarded to the subsequent encoder.

\section{PIVOT Methodology}
 
\definecolor{darkgoldenrod}{rgb}{0.72, 0.53, 0.04}
\begin{figure*}
    \centering
    \includegraphics[width=0.7\textwidth]{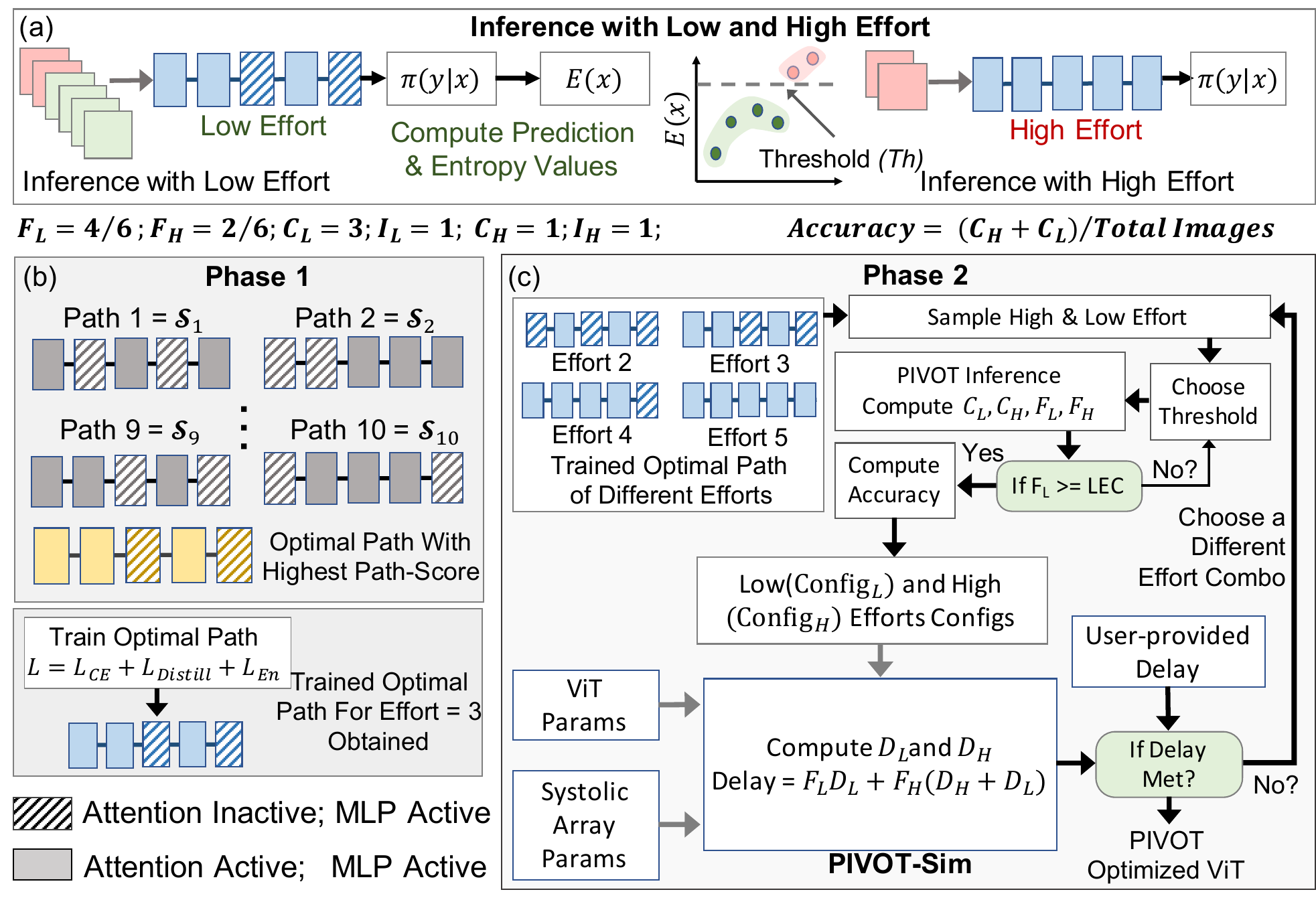}
    \caption{Figure showing (a) Input difficulty-aware inference procedure with PIVOT (b) PIVOT's Phase 1 (b) Phase2 Methodology. $LEC$ denotes the user-provided low effort constraint which implies the fraction of inputs that must be classified by the low effort ViT. For PIVOT-Sim, ViT params include embedding dim size, mlp ratio etc. and systolic array params include array size, dataflow, etc.} 
    \label{fig:inf_phase1_phase2}
\end{figure*}
\subsection{PIVOT Inference with Low and High Efforts}
\label{sec:inference}
During PIVOT's inference, we use the entropy metric to determine the number of attentions required to classify an input \cite{li2023input}. The entropy, $E(x)$, for an input $x$ (belonging to a dataset with $K$ classes) is calculated using Equation \ref{eq:entropy}. Here, $\pi(\bm{y}|\bm{x})$ is the logit output of the ViT. The term $1/\log K$ normalizes the final entropy to $(0,1]$. 
\begin{equation}
    E(\bm{x}) = -\frac{1}{\log K}\sum_{i=1}^K  \pi(\bm{y}_i|\bm{x})\log \pi(\bm{y}_i|\bm{x}).
    \label{eq:entropy}
\end{equation}
The entropy measures the confidence of prediction. For example, if all classes have an equal probability of $\frac{1}{K}$, the entropy value will be 1, implying uncertainty in the prediction. Whereas, if one class's prediction probability reaches 1 while the other classes attain 0 probability, the entropy reaches 0 implying confident prediction. 

As shown in Fig. \ref{fig:inf_phase1_phase2}a, during inference, PIVOT uses a combination of two efforts: 1) {Low Effort} and 2) {High Effort}. Here, \textit{Effort} is defined as the number of active attention modules (attentions that are not skipped) in the ViT. First, all inputs are inferred with the low effort resulting in the logit outputs ($\pi(\bm{y}|\bm{x})$) and the entropy values ($E(x)$). For inputs with entropy values lower than the threshold ($Th$), the $\pi(\bm{y}|\bm{x})$ from the low effort ViT are used for class prediction. For inputs with $E(x) > Th$, an additional inference is performed with high effort and then, all inputs are inevitably classified. In Fig. \ref{fig:inf_phase1_phase2}a, $F_L$ and $F_H$ are defined as the fraction of inputs classified by low ($E(x)<Th$) and high effort ($E(x)>Th$), respectively. Additionally, the number of inputs correctly (incorrectly) classified with low and high efforts are denoted as $C_L$ ($I_L$) and $C_H$ ($I_H$), respectively. The $C_L$ and $C_H$ values are used to compute the accuracy. 

\textbf{Re-computation Overhead: }During inference, some of the inputs that are unclassified with the low-effort are re-inferred with the high effort which entails re-computation overhead that needs to be managed to obtain a tradeoff between accuracy vs. efficiency. 



\subsection{PIVOT Phase1: Optimal Path Selection }
\label{sec:phase1}



PIVOT uses a two-phase hardware-in-the-loop search to design the multi-effort ViT. In Phase1, we select the optimal path for different efforts for a given ViT. Each effort contains multiple \textit{Paths}. For example, as shown in Fig. \ref{fig:inf_phase1_phase2}b, a ViT with 5 encoders and \textit{Effort}=3 entails ${5\choose 3} = 10$ possible {paths}. Here, a \textit{Path} is uniquely defined by the position of encoders with active and inactive attention modules. Having large number of paths for each effort increases the search space size in Phase2. Therefore, we define a \textit{Path-Score} (shown in Algorithm \ref{alg:path_score}) metric to single-out the \textit{Optimal Path} (shown in \textcolor{darkgoldenrod}{yellow}) corresponding to each {effort}. The {path} with the highest \textit{Path-Score} ($\mathcal{S}$) is chosen as the \textit{Optimal Path} and trained with the loss function shown in Fig \ref{fig:inf_phase1_phase2}b. The loss function contains cross-entropy loss ${L_{CE}}$, and the distillation loss ${L}_{Distill}$ between the final layer features of the teacher and student ViT. The $L_{CE}$ and $L_{Distill}$ are commonly used in prior works to train high performance ViTs \cite{touvron2021training, jiang2021all}. In PIVOT, to improve the prediction confidence, we add the regularization term $L_{En}$ that lowers the entropy for the correctly classified inputs. $L_{En}$ is the mean of the entropy values for the correctly classified inputs. Lowering the entropy ensures increased confident classifications with low efforts and thereby improves the inference efficiency.

\begin{figure}
    \centering
    \includegraphics[width=1\linewidth]{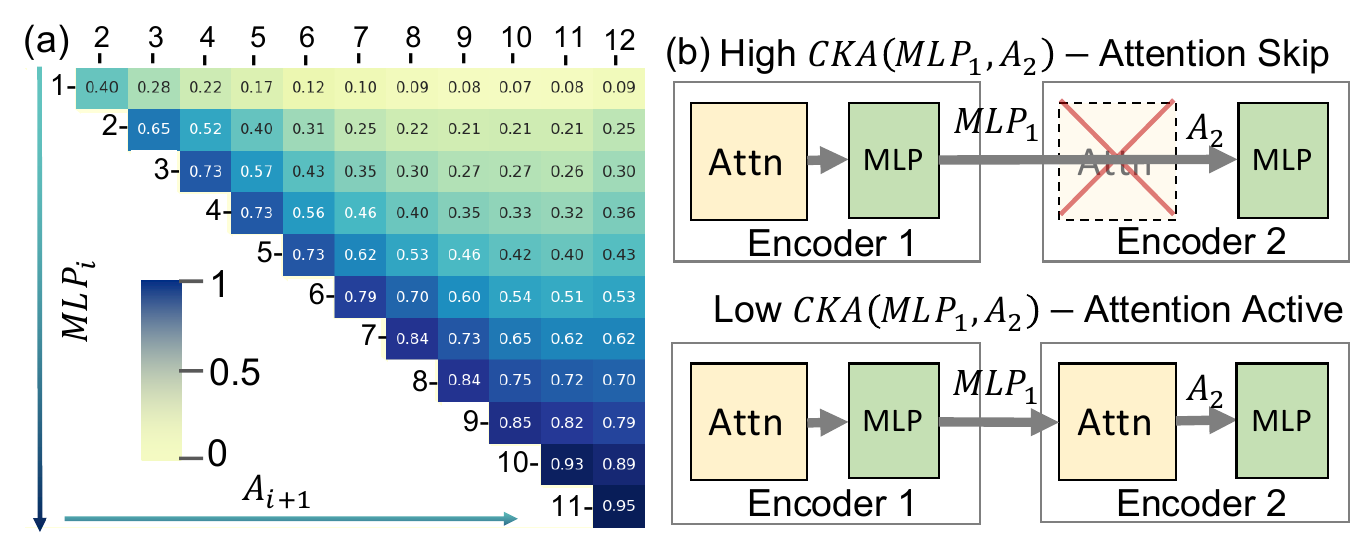}
    \caption{(a) $CKA ~Matrix$ computed between the MLP output of $Encoder_i$ ($MLP_i$) and Attention output of $Encoder_{i+1}$ ($A_{i+1}$) for the DeiT-S ViT (b) Higher $CKA(MLP_i,A_{i+1})$ suggests data redundancy and the attention can be skipped.} 
    \label{fig:cka_mat_demo}
\end{figure}

\textbf{CKA Matrix} Fig. \ref{fig:cka_mat_demo}a shows the center kernel alignment matrix ($CKA ~Matrix$) comprising of the CKA values computed between MLP outputs ($MLP_i$) and attention outputs ($A_{i+1}$) of ViT encoders $Encoder_i$ and $Encoder_{i+1}$, respectively. CKA measures the similarity between two matrices \cite{cortes2012algorithms}. A high $CKA(MLP_i,A_{i+1})$ value implies high similarity in $MLP_i$ and $A_{i+1}$ outputs, thus suggesting that output $MLP_i$ can be directly forwarded to $MLP_{i+1}$ by skipping $A_{i+1}$ as shown in Fig. \ref{fig:cka_mat_demo}b (top). Contrarily, for a low $CKA(MLP_i,A_{i+1})$ value, the attention cannot be skipped as shown in Fig. \ref{fig:cka_mat_demo}b (bottom). 

\begin{algorithm}
\KwIn{Effort Configuration ($Config$), \#Encoders in ViT ($D$), $CKA ~Matrix$.} 
\KwOut{Path-Score $(\mathcal{S})$}
\caption{Path-Score Computation Algorithm}
\label{alg:path_score}
$\mathcal{S} = 0$\;
\For{$i$ $\epsilon$ $Config$} {
    \For{$j$ $\epsilon$ ($i+1$, $D$)} {
        
        \If{($A_j$ is $Inactive$)}{
            $\mathcal{S}$ = $\mathcal{S} + CKA ~Matrix(i, j)$\;
        }
        \Else{
             break\;
        }
    }
}
\end{algorithm}
\begin{figure*}
    \centering
    \includegraphics[width=0.8\textwidth]{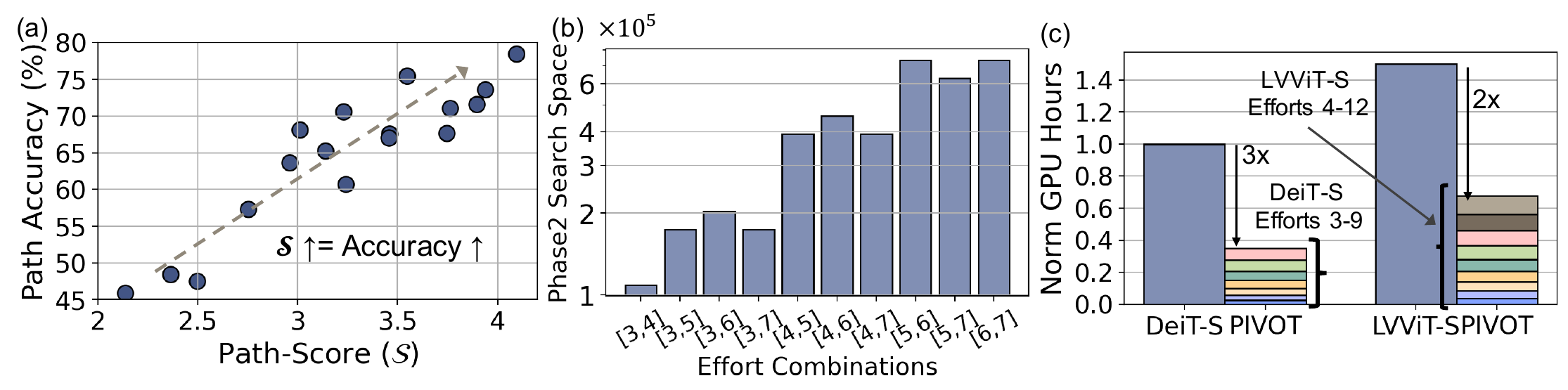}
    \caption{(a) Path Accuracy vs. \textit{Path-Score} ($\mathcal{S}$) corresponding to Effort = 6 for DeiT-S ViT. (b) Design space size if random search is performed in Phase2, without selecting optimal path for each effort in Phase1 (size normalized to PIVOT's design space size) (c) GPU hours for training DeiT-S, LVViT-S and PIVOT Efforts (normalized to GPU hours required for training DeiT-S from scratch).}
    \label{fig:path_ds_gpuhr}
\end{figure*}
\textbf{Path-Score} ($\mathcal{S}$): Algorithm \ref{alg:path_score} shows the methodology to compute $\mathcal{S}$. Algorithm \ref{alg:path_score} requires the $CKA ~Matrix$ (shown in Fig. \ref{fig:cka_mat_demo}a) and the effort configuration ($Config$), containing encoder locations with active and inactive attention. The $CKA ~Matrix$ is generated for a small batch of 256 images. For a given $Config$, $\mathcal{S}$ is computed by summing up the CKA values between the MLP outputs ($MLP$) of the encoders with active attention and the attention outputs ($A$) of the encoders with inactive attention. For example, $\mathcal{S}$ for $Config=$ [1,2,\textcolor{cyan}{3},\textcolor{cyan}{4},5,6,7,8,\textcolor{cyan}{9},\textcolor{cyan}{10},11,12], where encoder indices of inactive attentions are denoted by \textcolor{cyan}{cyan} can be computed as $CKA[MLP_2,A_3]+CKA[MLP_2,A_4]+CKA[MLP_8,A_9]+CKA[MLP_8,A{10}]$.
A high $\mathcal{S}$ signifies that the path contains highly redundant attentions that can be easily pruned out. Fig. \ref{fig:path_ds_gpuhr}a shows the positive correlation between $\mathcal{S}$ and path accuracy. As high $S$ paths ensure pruning the most redundant attention blocks, they attain higher accuracy.

\subsection{PIVOT Phase2: Selecting Optimal Effort Combinations}
In Phase2, given a set of efforts with optimal paths (shown in \textcolor{blue}{blue} in Fig. \ref{fig:inf_phase1_phase2}c), PIVOT determines the right effort combination to achieve optimal accuracy while meeting the user-provided delay requirement. 1) First, we start with a pair of low and high efforts (say, Effort 9 and Effort 12). 2) Next, the threshold values $Th$ for the low effort inference is chosen. The $Th$ values are iterated in an incremental manner. 3) A small batch of data (randomly sampled batch of 256 images from the training set) is inferred with the low and high efforts. This generates the $C_L$, $C_H$, $F_L$ and $F_H$ values. 4) Following this, the accuracy calculator uses $C_L$ and $C_H$ to compute the accuracy (Fig. \ref{fig:inf_phase1_phase2}a). The thresholds are iterated until the condition $F_L \geq LEC$ is met. Higher $LEC$ value ensures more inputs classified by the low effort ViT. 5) The low ($Config_L$), high ($Config_H$) effort configurations, $F_L$ and $F_H$ values are passed to the PIVOT-Sim framework for delay computation. The PIVOT-Sim platform first computes the delays of low and high efforts ($D_L$ and $D_H$, respectively) using $Config_L$, $Config_H$, ViT and systolic array parameters (Refer Section \ref{sec:pivot_sim}). Then, it computes the delay of the effort combination using $D_L$, $D_H$, $F_L$ and $F_H$. If the delay lies within 5\% of the user-provided delay constraint, the optimal effort combination is obtained. If the delay constraint is not met, a new effort combination (say, Effort 6 and Effort 9) is selected. In order to achieve high accuracy, the sampling starts with efforts containing maximum active attentions. In each iteration, a smaller effort combination is sampled than the previous iteration until the desired delay is obtained. 

\textbf{Benefit of CKA Score-based Optimal Path Selection} In Fig. \ref{fig:path_ds_gpuhr}b, we compare the Phase2 design space size of random and PIVOT-based search. Since PIVOT uses the \textit{Path-score} to single out the optimal path for each effort, there exists only one path for each effort combination. Whereas, Phase2 with random search entails multiple paths due to the absence of optimal path selection. For example, in random search as shown in Fig. \ref{fig:path_ds_gpuhr}b, effort combinations [3,6] can contain ${12\choose3} \times {12\choose 6} = 2.03\times 10^5$ possible paths for the DeiT-S ViT. For the DeiT-S ViT Phase2 with random search, the search space size is $\sim10^5\times$ higher than PIVOT's search space size. 

\textbf{GPU Hours for Training all Efforts:}  Fig. \ref{fig:path_ds_gpuhr}c shows that the combined GPU hours required for training all efforts (see Section \ref{sec:exp_setup}) for DeiT-S (LVViT-S) ViTs in PIVOT is 3$\times$ (2$\times$) less compared to training the DeiT-S (LVViT-S) ViT from scratch. This is because, the training time reduces with reduction in the ViT effort. 

\subsection{PIVOT-Sim Platform}
\label{sec:pivot_sim}

\begin{figure}[h!]
    \centering
    \includegraphics[width=1\linewidth]{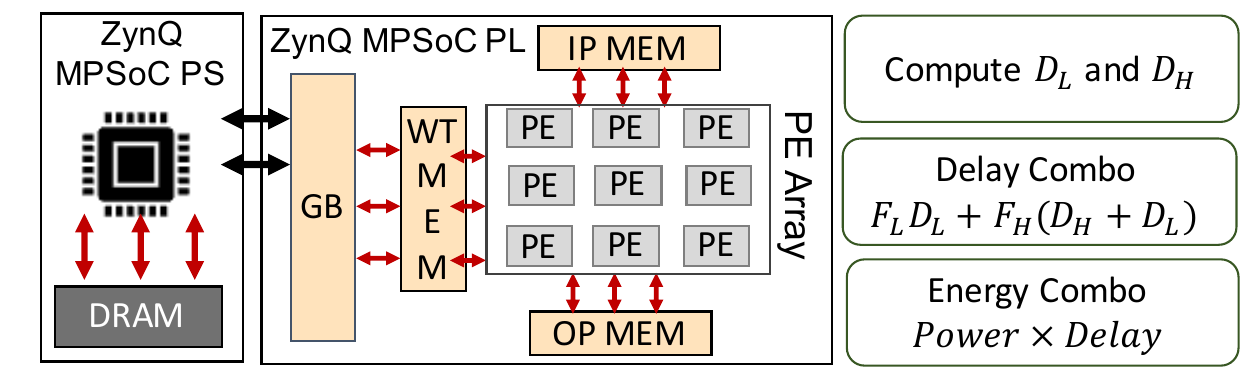}
    \caption{Figure showing the PIVOT-Sim Platform.} 
    \label{fig:pivot_sim}
\end{figure}
Fig. \ref{fig:pivot_sim} shows the overall architecture of the PIVOT-Sim platform. PIVOT-Sim performs cycle-accurate delay estimation for a given ViT effort mapped on a Xilinx ZCU102 MPSoC FPGA-based systolic array accelerator. Like the ZynQ MPSOC FPGA, PIVOT-Sim contains two systems: 1) ZynQ MPSoC Processing System (PS) and 2) ZynQ MPSoC programmable logic (PL). All the linear matrix multiplication layers (QKV, QK$^T$, SMxV, Proj and MLP) are executed in the PL-implemented systolic accelerator. Inputs and weights are first loaded from the PS DRAM to the global SRAM buffer (GB) in the PL. Then the weights and inputs are fetched from GB to the Weight SRAM (WTMEM) and Input SRAM (IPMEM), respectively. Then, the weights from the WTMEM are loaded in to the PE array in a streaming fashion following which, the inputs are fetched from the IPMEM in a streaming fashion column by column. The multiply-and-accumulate (MAC) outputs are stored in the output SRAM (OPMEM). The outputs are pushed to the GB and finally to the DRAM. The non-linear operations such as softmax, entropy and GeLU are implemented using the ZynQ MPSoC PS. 

The PIVOT-Sim framework requires the ViT parameters (embedding dimension size, number of tokens, mlp ratio and attention head count) and systolic array parameters (array dimensions, dataflow, SRAM memory sizes, and the clock frequency) and the low (high) effort configurations $Config_L$ ($Config_H$) (discussed in Section \ref{sec:phase1}) to compute the low (high) effort delays $D_L$ ($D_H$). Additionally, it also computes the delay of low-high effort combination using the $F_L$, $F_H$, $D_L$ and $D_H$ values as shown in Fig. \ref{fig:pivot_sim}. The $D_L\times F_H$ term in the delay computation accounts for the re-computation overhead. The energy is obtained by multiplying the power with the delay of the effort combination. 

\textbf{Entropy Computation Overhead} We find that entropy computation (Equation \ref{eq:entropy}) in the ZynQ MPSoC PS takes 0.03ms per image which is $<0.05\%$ of the inference delay and thus, can be ignored. 

\section{Experiments and Results}
\subsection{Experimental Setup}
\label{sec:exp_setup}
\textbf{Datasets and ViTs:} We benchmark all our results on the standard Imagenet-1K dataset using state-of-the-art efficient ViTs such as DeiT \cite{touvron2021training} and LV-ViT \cite{jiang2021all}. \textbf{Baseline: }For all experiments, the baseline is a ViT model without any effort modulation \textit{i.e.,} all ViT attention modules will be activated irrespective of the input difficulty. \textbf{PIVOT-Optimized ViTs:} For ease of expression, throughout the text, we will refer to PIVOT-optimized DeiT-S and LVViT-S ViT as PVDS and PVLS, respectively.

\textbf{Traning Details:} In PIVOT, for the DeiT-S and LVViT-S ViTs, we create 7 (3, 4, 5, 6, 7, 8 and 9) and 9 (4, 5, 6, 7, 8, 9, 10, 11, 12) efforts, respectively. Each effort is finetuned for 30 epochs with the full training data. The ViTs are trained with 8-bit quantization. Training all the efforts is 3$\times$ (2$\times$) more efficient than training a DeiT-S (LVViT-S) ViT from scratch (see Fig. \ref{fig:path_ds_gpuhr}c). For training we use Pytorch 1.3.1 with a single Nvidia V100 GPU backend. 

\textbf{Hardware Evaluation:} All baselines and PIVOT-optimized ViTs (PVDS and PVLS) are evaluated using the PIVOT-Sim framework. The FPGA implementation parameters for PIVOT-Sim are shown in Table \ref{tab:hardware_resources}. The FPGA implementation requires 4566 LUTs, 20668 Registers, 48 Block RAMs and 2304 digital signal processing cores. 

\begin{table}[h!]
    \Huge
    \centering
    \resizebox{0.6\linewidth}{!}{\begin{tabular}{|c|c|c|c|} \hline
        \multicolumn{2}{|c|}{FPGA Board} & \multicolumn{2}{c|}{Xilinx ZCU102} \\ \hline
        \multicolumn{2}{|c|}{Global SRAM (GB) Size} & \multicolumn{2}{|c|}{16KB} \\ \hline
        \multicolumn{2}{|c|}{IPMEM, WTMEM, OPMEM} & \multicolumn{2}{|c|}{64Kb, 64Kb, 64Kb} \\  \hline
        \multicolumn{2}{|c|}{PE Array Size} & \multicolumn{2}{|c|}{64$\times$36} \\ \hline
        \multicolumn{2}{|c|}{Clock Frequency} & \multicolumn{2}{|c|}{125MHz} \\ \hline
        \multicolumn{2}{|c|}{Dataflow} & \multicolumn{2}{|c|}{Input Stationary} \\ \hline 
        \end{tabular}}
    \caption{Table showing the FPGA implementation parameters.} 
    \label{tab:hardware_resources}
\end{table}

\subsection{Results on DeiT-S and LVViT-S ViTs} 

\begin{table}[h!]
\Huge
\centering
\caption{Table comparing the performance of DeiT-S and PIVOT-optimized DeiT-S ViTs (PVDS-$N$) sampled at  delay=$N$.}
\resizebox{\linewidth}{!}{
\begin{tabular}{|c|c|c|c|c|c|c|}
\hline
\multirow{2}{*}{\textbf{Model}} & \multirow{2}{*}{\begin{tabular}{c}
     \textbf{Energy} \\ \textbf{(J)}
\end{tabular}} & 

\multirow{2}{*}{\begin{tabular}{c}
     \textbf{Delay} \\ \textbf{(ms)}
\end{tabular}} & 

\multirow{2}{*}{\begin{tabular}{c}
     \textbf{Power} \\ \textbf{(W)}
\end{tabular}} & 

\multirow{2}{*}{\begin{tabular}{c}
     \textbf{EDP} \\ \textbf{(J$\times$ms)}
\end{tabular}} & 

\multirow{2}{*}{\textbf{FPS/W}} & \multirow{2}{*}{\begin{tabular}{c}
     \textbf{Accuracy} \\ \textbf{(\%)}
\end{tabular}} \\ 

& & & & & & \\ \hline

{DeiT-S} & 0.47 & 59.66 & 7.92 & 28.19 & 2.14(1$\times$) & 79.8 \\ \hline 
PVDS-50 & 0.38 (1.23$\times$)  & 48.47 ($1.23\times$) & 7.92 & 16.21 (1.73$\times$) & 2.7(1.23$\times$) & 79.4 \\ \hline


PVDS-35 & 0.292 (1.62$\times$)  & 36.9 ($1.61\times$) & 7.92 & 10.5 (2.6$\times$) & 3.4(1.61$\times$) & 78.2\\ \hline

\end{tabular}} 
\label{table:deits-results}
\end{table}

\begin{table}[h!]
\Huge
\centering
\caption{Table comparing the performance of LVViT-S and PIVOT-optimized LVViT-S ViTs (PVLS-$N$) sampled at  delay=$N$. }
\resizebox{\linewidth}{!}{
\begin{tabular}{|c|c|c|c|c|c|c|}
\hline
\multirow{2}{*}{\textbf{Model}} & \multirow{2}{*}{\begin{tabular}{c}
     \textbf{Energy} \\ \textbf{(J)}
\end{tabular}} & 

\multirow{2}{*}{\begin{tabular}{c}
     \textbf{Delay} \\ \textbf{(ms)}
\end{tabular}} & 

\multirow{2}{*}{\begin{tabular}{c}
     \textbf{Power} \\ \textbf{(W)}
\end{tabular}} & 

\multirow{2}{*}{\begin{tabular}{c}
     \textbf{EDP} \\ \textbf{(J$\times$ms)}
\end{tabular}} & 

\multirow{2}{*}{\textbf{FPS/W}} & \multirow{2}{*}{\begin{tabular}{c}
     \textbf{Accuracy} \\ \textbf{(\%)}
\end{tabular}} \\ 

& & & & & & \\ \hline

{LVViT-S} & {0.63} & {79.55} & {7.92} &  {50.8} & 1.57(1$\times$) & {82.8} \\ \hline 


PVLS-50 & 0.410 (1.57$\times$)  & 50 ($1.6\times$) & 7.92 & 20.13 (2.7$\times$) & 2.51(1.6$\times$) & 82.6 \\ \hline

PVLS-35 & 0.312 (2.17$\times$) & 36.5 ($2.17\times$) & 7.92 & 10.57 (4.5$\times$) & 3.4(2.17$\times$) & 81.1 \\ \hline

\end{tabular}}
\label{table:lvvits-results}
\end{table}


Table \ref{table:deits-results} and Table \ref{table:lvvits-results} compare the delay, energy-delay-product (EDP), energy efficiency (FPS/W) and the accuracy of different PVDS and PVLS ViTs searched at different target delays lesser than the baseline. Evidently, as seen in Table \ref{table:deits-results} the PVDS-50 (PVLS-50) ViTs achieve 1.73$\times$ (2.7$\times$) EDP reduction, 1.23$\times$ (1.6$\times$) higher FPS/W with merely 0.4\% (0.2\%) accuracy reduction compared to the baseline DeiT-S (LVViT-S). At a slightly higher accuracy reduction of 1.6\% (1.7\%) the PVDS-35 (PVLS-35) yields 2.6$\times$ (4.5$\times$) lower EDP and 1.62$\times$ (2.17$\times$) higher FPS/W compared to baseline. 

\begin{figure}[h!]
    \centering
    \includegraphics[width=\linewidth]{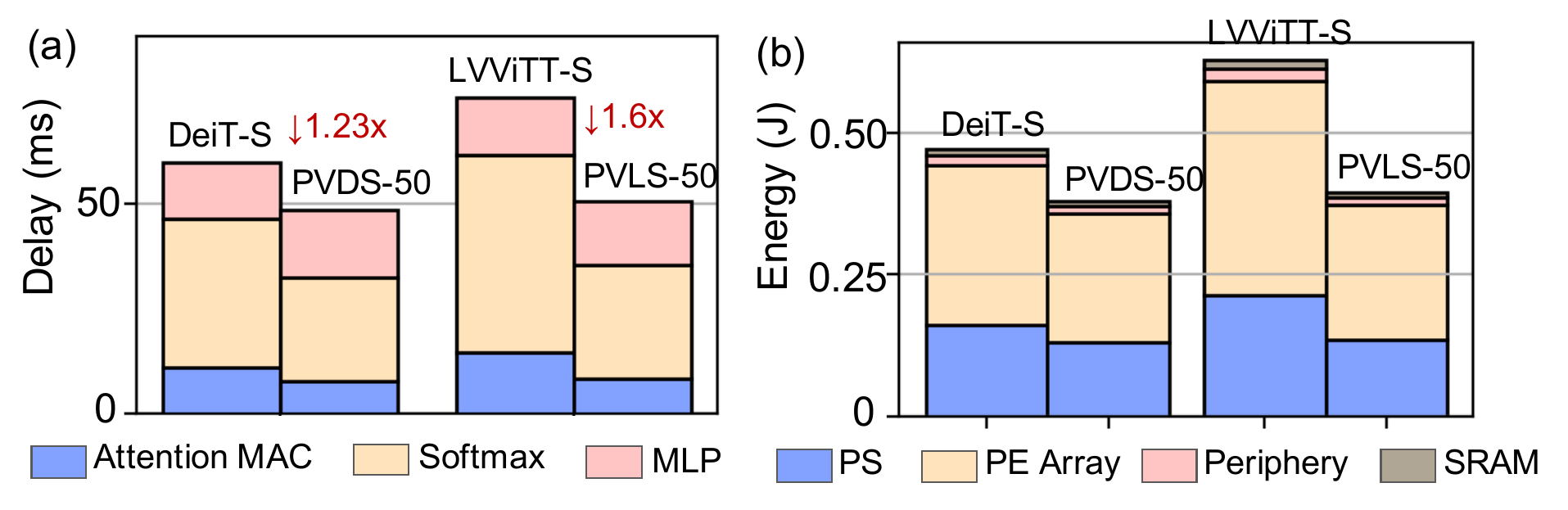}
    \caption{(a) Delay breakdown across encoder modules for different ViTs (b) Energy breakdown across the PE Array, Periphery and SRAM (part of the ZynQ MPSoC PL) and the PS (ZynQ MPSoC PS).} 
    \label{fig:delay_dist_net}
\end{figure}
Fig. \ref{fig:delay_dist_net}a shows the delay distributions across the Attention MAC (QKV, QK$^T$, (SMxV) and Proj), Softmax and MLP modules (refer Fig. \ref{fig:flop_red_latency_ov}a). Interestingly, the softmax module consumes 60\% (63\%) of the overall delay in the DeiT-S (LVViT-S) ViTs. With PIVOT, the softmax overhead reduces to 43\% (48\%) for the PVDS-50 (PVLS-50) ViTs. Similarly, the Attention MAC overhead reduces to 13\% (14\%) in the PVDS-50 (PVLS-50) ViTs compared to 18\% (19\%) in DeiT-S (LVViT-S) ViTs. Note, since PIVOT does not skip MLP modules, the delay overhead of MLP in PVDS-50 (PVLS-50) increase by 21\% (19\%) compared to the baselines due to the re-computation overhead (refer Section \ref{sec:inference}). However, due to high delay reduction in softmax and attention MAC modules, PIVOT achieves an overall delay reduction. 

\textbf{Energy Reduction across FPGA Resources: }As seen in Fig. \ref{fig:delay_dist_net}b, delay reduction in PVDS-50 and PVLS-50 ViTs lead to an energy reduction across the ZynQ MPSoC PS and PL systems. PVDS-50 and PVLS-50 ViTs achieve around 2$\times$ energy reduction in the PS and 1.6$\times$, 1.7$\times$ and 1.8$\times$ energy reduction in the PE-Array, SRAM memories and peripheral circuits, respectively implemented on the ZynQ MPSoC PL (See Section. \ref{sec:pivot_sim}). The peripheral circuits (periphery) include PS-PL interconnects, reset and memory controllers. 



\subsection{Comparison with Prior Works}
\begin{table}[h!]
\Huge
\centering
\caption{Performance comparison of ViTCOD \cite{you2023vitcod}, HeatViT \cite{dong2023heatvit} and PVDS-50.}
\label{tab:comp_prior_works}
\resizebox{0.8\linewidth}{!}{
\begin{tabular}{|c|c|c|c|}
\hline

{Work} & {ViTCOD} \cite{you2023vitcod} & {HeatViT} \cite{dong2023heatvit} & \textbf{PIVOT (Ours)} \\ \hline
ViT Backbone & DeiT-S & DeiT-S & \textbf{DeiT-S} \\ \hline
Effort Modulation & Constant & Constant & \textbf{Input-aware} \\ \hline
Prediction & Norm & Head & \textbf{Entropy} \\
Mechanism & Score & Level & \textbf{Metric} \\ \hline
Quantization & 8-bits & 8-bits & \textbf{8-bits} \\ \hline
Accuracy & 78.1\% & 79.1\% & \textbf{79.4\%} \\ \hline
GPP Compatible & $\times$ & $\times$ & \textbf{\checkmark} \\ \hline


\end{tabular}} 
\end{table}

Table \ref{tab:comp_prior_works} performs a holistic comparison between PIVOT and prior state-of-the-art algorithm-hardware co-design frameworks \cite{dong2023heatvit, you2023vitcod}.
Soft token pruning in HeatViT \cite{dong2023heatvit} achieves a high token pruning ratio of 40\%, 74\% and 87\% in encoders 4-6, 7-9, and 10-12, respectively, while achieving 79.1\% accuracy. ViTCOD \cite{you2023vitcod} achieves 90\% attention sparsity ratio at 78.1\% accuracy. \textbf{Accuracy advantage in PIVOT:} HeatViT \cite{dong2023heatvit} and ViTCOD \cite{you2023vitcod} do not modulate their efforts based on the input difficulty (token and attention sparsity ratios remain constant for all inputs). Therefore, at high token and attention pruning ratios, the accuracy suffers as difficult images are wrongly classified. Whereas, due to input-awareness, PIVOT (PVDS-50) achieves the highest accuracy of 79.4\%. 

\begin{figure}
    \centering
    \includegraphics[width=\linewidth]{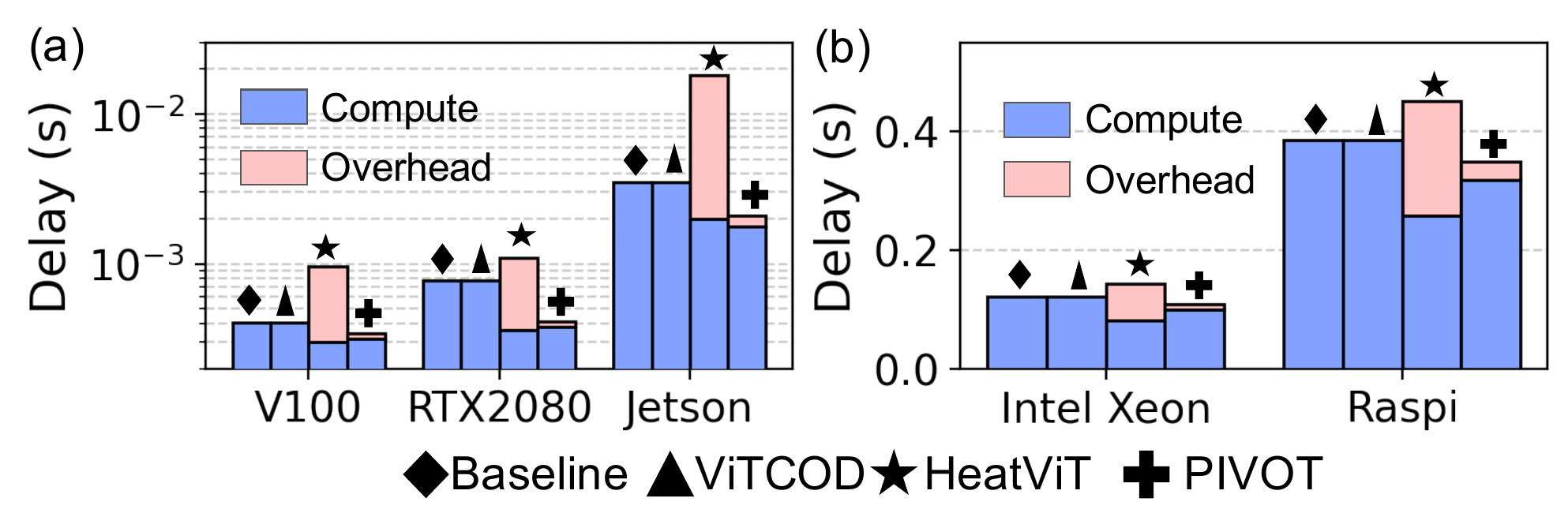}
    \caption{Compute and overhead delay breakdowns for DeiT-S baseline, HeatViT \cite{dong2023heatvit}, ViTCOD \cite{you2023vitcod} and PIVOT (PVDS-50) across (a) Nvidia V100, NVidia RTX2080ti and Nvidia Jetson Orin Nano (b) Intel Xeon and Raspberry Pi 4. } 
    \label{fig:prior_works}
\end{figure}
\textbf{Evaluation on GPPs: }As HeatViT \cite{dong2023heatvit} and ViTCOD \cite{you2023vitcod} require special hardware support for efficient implementation, we perform the delay comparison on GPPs such as CPUs- Intel Xeon, Raspberry Pi, and GPUs- Nvidia V100, Nvidia RTX2080ti and Nvidia Jetson Orin Nano for a fair comparison. As seen in Fig. \ref{fig:prior_works}a and Fig. \ref{fig:prior_works}b, the PIVOT (PVDS-50) achieves around 1.2-1.5$\times$ lower delay compared to the baseline across all GPPs. Since ViTCOD requires sparse matrix multiplication support, the delay on GPP is similar to the baseline. Due to hefty predictor networks and token packaging modules for soft token pruning, HeatViT \cite{dong2023heatvit} entails significant delay overhead when implemented on GPPs. PIVOT is general purpose and entails a small overhead of 6\% in the delay. This delay is majorly contributed by the re-computation overhead. The contribution of entropy computation (Equation \ref{eq:entropy}) is negligibly small ($<0.05\%$). 

\subsection{Analysis with $LEC$ Constraints}
\begin{figure}
    \centering
    \includegraphics[width=1\linewidth]{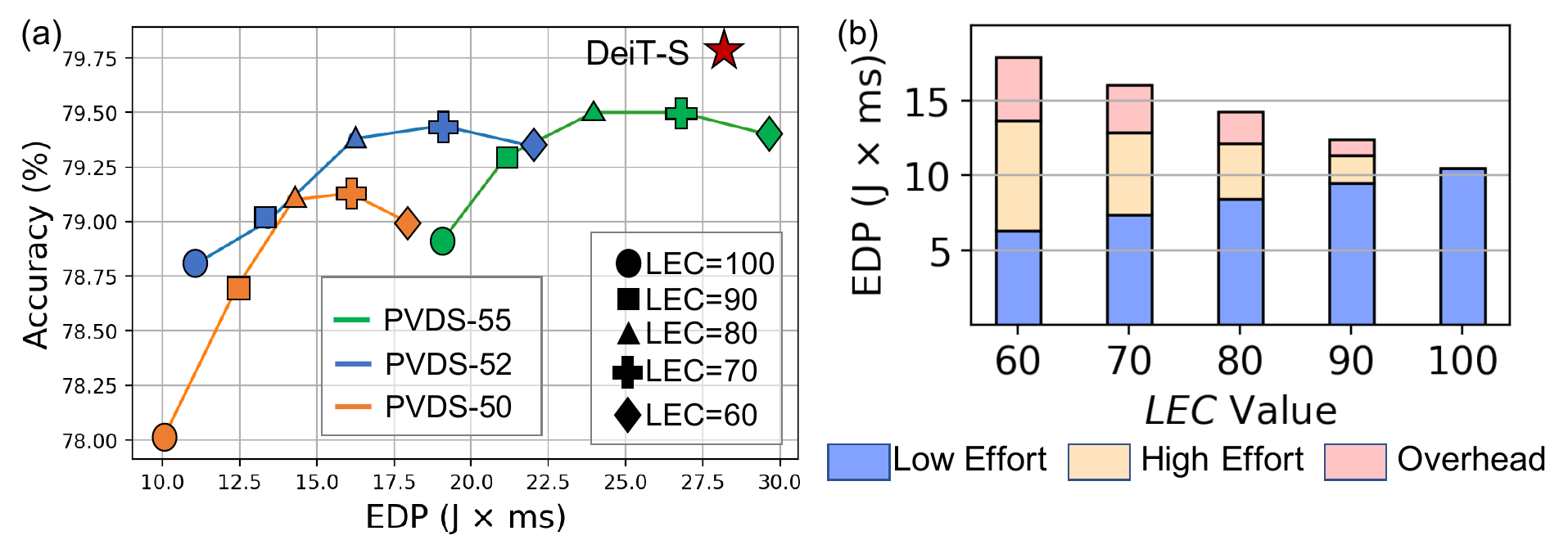}
    \caption{Figure analysing the effect of different $LEC$ on the EDP and accuracy for different effort combinations. (b) EDP distribution between the low effort, high effort and the re-computation overhead (Overhead) for the PVDS-50 ViT.} 
    \label{fig:EDP_Acc_Dist}
\end{figure}
From Fig. \ref{fig:EDP_Acc_Dist} we find that $LEC=70$ and $LEC=80$ attain the best EDP and accuracy tradeoff across different PVDS ViTs. At low $LEC=60$, the EDP is high as merely 60\% of the inputs are classified by the low effort. Additionally, $LEC=90$ entails 90\% of the inference with low effort but this leads to a significant accuracy degradation.

The EDP is contributed by the low effort and high effort inference, and the re-computation overhead (Section \ref{sec:inference}). At low $LEC$ values, both high-effort and re-computation EDPs are high while the low effort EDP is less. As the $LEC$ value increases, the low effort EDP increases marginally while the high effort and re-computation EDP reduce significantly leading to overall low EDPs.

\textbf{Need for Input difficulty awareness} As seen in Fig. \ref{fig:EDP_Acc_Dist}a for $LEC=100$, all inputs are inferred by the low effort. This leads to low EDP at the cost of accuracy since the efforts are not modulated for difficult inputs. Therefore, PIVOT's input-aware effort modulation achieves optimal accuracy-efficiency tradeoffs.  

\subsection{Efforts Combinations for Different Delays}
\begin{figure}[h!]
    \centering
    \includegraphics[width=0.7\linewidth]{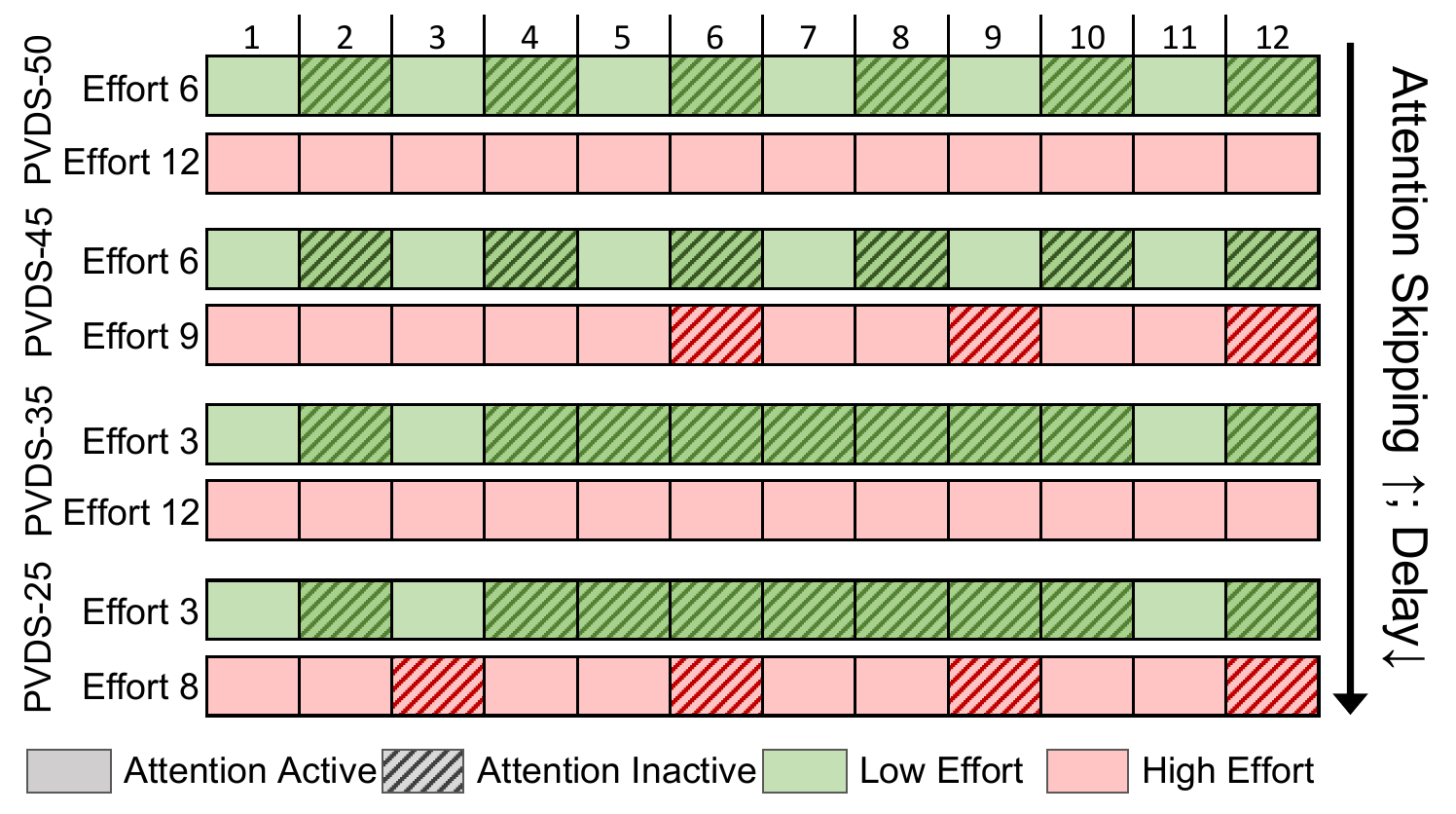}
    \caption{Different PVDS ViTs sampled by PIVOT at different delay constraints.}
    \label{fig:pv_nets}
\end{figure}

As seen in Fig. \ref{fig:pv_nets}, reduction in the delay requirement lowers the number of active attentions in the ViT. The efforts shown here represent the optimal path with the highest \textit{Path-score} for each effort. Interestingly, we observe that across all efforts, attentions skipping is preferred in the deeper layers as the $CKA(MLP, A)$ value is higher in the latter layers.

\section{Conclusion}
PIVOT motivates ViT attention optimization in an input difficulty-aware manner. PIVOT's input-awareness yields 0.4\%-1.3\% higher accuracy compared to prior token pruning and attention sparsification works. Unlike prior works, PIVOT is GPP compatible and yields 1.2-1.5$\times$ higher throughput compared to baseline ViT across different CPU/GPU platforms. Additionally, PIVOT-Sim- an end-to-end open source FPGA-based evaluation platform is developed that will motivate future ViT-hardware co-optimization works. 

\section*{Acknowledgement}
This work was supported in part by CoCoSys, a JUMP2.0 center sponsored by DARPA and SRC, the National Science Foundation (CAREER Award, Grant \#2312366, Grant \#2318152), and the DoE MMICC center SEA-CROGS (Award \#DE-SC0023198) 
\bibliographystyle{ACM-Reference-Format}
\bibliography{PIVOT.bib}


\begin{thebibliography}{17}


\ifx \showCODEN    \undefined \def \showCODEN     #1{\unskip}     \fi
\ifx \showDOI      \undefined \def \showDOI       #1{#1}\fi
\ifx \showISBNx    \undefined \def \showISBNx     #1{\unskip}     \fi
\ifx \showISBNxiii \undefined \def \showISBNxiii  #1{\unskip}     \fi
\ifx \showISSN     \undefined \def \showISSN      #1{\unskip}     \fi
\ifx \showLCCN     \undefined \def \showLCCN      #1{\unskip}     \fi
\ifx \shownote     \undefined \def \shownote      #1{#1}          \fi
\ifx \showarticletitle \undefined \def \showarticletitle #1{#1}   \fi
\ifx \showURL      \undefined \def \showURL       {\relax}        \fi
\providecommand\bibfield[2]{#2}
\providecommand\bibinfo[2]{#2}
\providecommand\natexlab[1]{#1}
\providecommand\showeprint[2][]{arXiv:#2}

\bibitem[Bhattacharjee et~al\mbox{.}(2022)]%
        {bhattacharjee2022mime}
\bibfield{author}{\bibinfo{person}{Bhattacharjee} {et~al\mbox{.}}} \bibinfo{year}{2022}\natexlab{}.
\newblock \showarticletitle{MIME: adapting a single neural network for multi-task inference with memory-efficient dynamic pruning}. In \bibinfo{booktitle}{\emph{Proceedings of the 59th ACM/IEEE Design Automation Conference}}. \bibinfo{pages}{499--504}.
\newblock


\bibitem[Cortes et~al\mbox{.}(2012)]%
        {cortes2012algorithms}
\bibfield{author}{\bibinfo{person}{Cortes} {et~al\mbox{.}}} \bibinfo{year}{2012}\natexlab{}.
\newblock \showarticletitle{Algorithms for learning kernels based on centered alignment}.
\newblock \bibinfo{journal}{\emph{The Journal of Machine Learning Research}} \bibinfo{volume}{13}, \bibinfo{number}{1} (\bibinfo{year}{2012}), \bibinfo{pages}{795--828}.
\newblock


\bibitem[Dehghani et~al\mbox{.}(2023)]%
        {dehghani2023scaling}
\bibfield{author}{\bibinfo{person}{Dehghani} {et~al\mbox{.}}} \bibinfo{year}{2023}\natexlab{}.
\newblock \showarticletitle{Scaling vision transformers to 22 billion parameters}. In \bibinfo{booktitle}{\emph{International Conference on Machine Learning}}. PMLR, \bibinfo{pages}{7480--7512}.
\newblock


\bibitem[Dong et~al\mbox{.}(2023)]%
        {dong2023heatvit}
\bibfield{author}{\bibinfo{person}{Dong} {et~al\mbox{.}}} \bibinfo{year}{2023}\natexlab{}.
\newblock \showarticletitle{Heatvit: Hardware-efficient adaptive token pruning for vision transformers}. In \bibinfo{booktitle}{\emph{2023 IEEE International Symposium on High-Performance Computer Architecture (HPCA)}}. IEEE, \bibinfo{pages}{442--455}.
\newblock


\bibitem[Dosovitskiy et~al\mbox{.}(2020)]%
        {dosovitskiy2020image}
\bibfield{author}{\bibinfo{person}{Dosovitskiy} {et~al\mbox{.}}} \bibinfo{year}{2020}\natexlab{}.
\newblock \showarticletitle{An image is worth 16x16 words: Transformers for image recognition at scale}.
\newblock \bibinfo{journal}{\emph{arXiv preprint arXiv:2010.11929}} (\bibinfo{year}{2020}).
\newblock


\bibitem[Han et~al\mbox{.}(2022)]%
        {han2022survey}
\bibfield{author}{\bibinfo{person}{Han} {et~al\mbox{.}}} \bibinfo{year}{2022}\natexlab{}.
\newblock \showarticletitle{A survey on vision transformer}.
\newblock \bibinfo{journal}{\emph{IEEE transactions on pattern analysis and machine intelligence}} \bibinfo{volume}{45}, \bibinfo{number}{1} (\bibinfo{year}{2022}), \bibinfo{pages}{87--110}.
\newblock


\bibitem[Jiang et~al\mbox{.}(2021)]%
        {jiang2021all}
\bibfield{author}{\bibinfo{person}{Jiang} {et~al\mbox{.}}} \bibinfo{year}{2021}\natexlab{}.
\newblock \showarticletitle{All tokens matter: Token labeling for training better vision transformers}.
\newblock \bibinfo{journal}{\emph{Advances in neural information processing systems}}  \bibinfo{volume}{34} (\bibinfo{year}{2021}), \bibinfo{pages}{18590--18602}.
\newblock


\bibitem[Kim et~al\mbox{.}(2021)]%
        {kim2021rethinking}
\bibfield{author}{\bibinfo{person}{Kim} {et~al\mbox{.}}} \bibinfo{year}{2021}\natexlab{}.
\newblock \showarticletitle{Rethinking the self-attention in vision transformers}. In \bibinfo{booktitle}{\emph{Proceedings of the IEEE/CVF Conference on Computer Vision and Pattern Recognition}}. \bibinfo{pages}{3071--3075}.
\newblock


\bibitem[Li et~al\mbox{.}(2023)]%
        {li2023input}
\bibfield{author}{\bibinfo{person}{Li} {et~al\mbox{.}}} \bibinfo{year}{2023}\natexlab{}.
\newblock \showarticletitle{Input-aware dynamic timestep spiking neural networks for efficient in-memory computing}. In \bibinfo{booktitle}{\emph{2023 60th ACM/IEEE Design Automation Conference (DAC)}}. IEEE, \bibinfo{pages}{1--6}.
\newblock


\bibitem[Panda et~al\mbox{.}(2016)]%
        {panda2016conditional}
\bibfield{author}{\bibinfo{person}{Panda} {et~al\mbox{.}}} \bibinfo{year}{2016}\natexlab{}.
\newblock \showarticletitle{Conditional deep learning for energy-efficient and enhanced pattern recognition}. In \bibinfo{booktitle}{\emph{2016 Design, Automation \& Test in Europe Conference \& Exhibition (DATE)}}. IEEE, \bibinfo{pages}{475--480}.
\newblock


\bibitem[Rao et~al\mbox{.}(2021)]%
        {rao2021dynamicvit}
\bibfield{author}{\bibinfo{person}{Rao} {et~al\mbox{.}}} \bibinfo{year}{2021}\natexlab{}.
\newblock \showarticletitle{Dynamicvit: Efficient vision transformers with dynamic token sparsification}.
\newblock \bibinfo{journal}{\emph{Advances in neural information processing systems}}  \bibinfo{volume}{34} (\bibinfo{year}{2021}), \bibinfo{pages}{13937--13949}.
\newblock


\bibitem[Samajdar et~al\mbox{.}(2018)]%
        {samajdar2018scale}
\bibfield{author}{\bibinfo{person}{Samajdar} {et~al\mbox{.}}} \bibinfo{year}{2018}\natexlab{}.
\newblock \showarticletitle{Scale-sim: Systolic cnn accelerator simulator}.
\newblock \bibinfo{journal}{\emph{arXiv preprint arXiv:1811.02883}} (\bibinfo{year}{2018}).
\newblock


\bibitem[Stevens et~al\mbox{.}(2021)]%
        {stevens2021softermax}
\bibfield{author}{\bibinfo{person}{Stevens} {et~al\mbox{.}}} \bibinfo{year}{2021}\natexlab{}.
\newblock \showarticletitle{Softermax: Hardware/software co-design of an efficient softmax for transformers}. In \bibinfo{booktitle}{\emph{2021 58th ACM/IEEE Design Automation Conference (DAC)}}. IEEE, \bibinfo{pages}{469--474}.
\newblock


\bibitem[Touvron et~al\mbox{.}(2021)]%
        {touvron2021training}
\bibfield{author}{\bibinfo{person}{Touvron} {et~al\mbox{.}}} \bibinfo{year}{2021}\natexlab{}.
\newblock \showarticletitle{Training data-efficient image transformers \& distillation through attention}. In \bibinfo{booktitle}{\emph{International conference on machine learning}}. PMLR, \bibinfo{pages}{10347--10357}.
\newblock


\bibitem[Wang et~al\mbox{.}(2021)]%
        {wang2021spatten}
\bibfield{author}{\bibinfo{person}{Wang} {et~al\mbox{.}}} \bibinfo{year}{2021}\natexlab{}.
\newblock \showarticletitle{Spatten: Efficient sparse attention architecture with cascade token and head pruning}. In \bibinfo{booktitle}{\emph{2021 IEEE International Symposium on High-Performance Computer Architecture (HPCA)}}. IEEE, \bibinfo{pages}{97--110}.
\newblock


\bibitem[Wu et~al\mbox{.}(2018)]%
        {wu2018blockdrop}
\bibfield{author}{\bibinfo{person}{Wu} {et~al\mbox{.}}} \bibinfo{year}{2018}\natexlab{}.
\newblock \showarticletitle{Blockdrop: Dynamic inference paths in residual networks}. In \bibinfo{booktitle}{\emph{Proceedings of the IEEE conference on computer vision and pattern recognition}}. \bibinfo{pages}{8817--8826}.
\newblock


\bibitem[You et~al\mbox{.}(2023)]%
        {you2023vitcod}
\bibfield{author}{\bibinfo{person}{You} {et~al\mbox{.}}} \bibinfo{year}{2023}\natexlab{}.
\newblock \showarticletitle{Vitcod: Vision transformer acceleration via dedicated algorithm and accelerator co-design}. In \bibinfo{booktitle}{\emph{2023 IEEE International Symposium on High-Performance Computer Architecture (HPCA)}}. IEEE, \bibinfo{pages}{273--286}.
\newblock


\end{thebibliography}










\end{document}